\documentclass[useAMS,usenatbib]{mn2e}
\usepackage{epsfig}
\usepackage{amssymb}
\usepackage{upgreek}
\usepackage{multirow}
\usepackage{enumerate}

\newcommand\ion[2]{#1\,{\scshape{#2}}}

\title[Reddening of NGC~5548] {Estimating reddening of the continuum and broad-line region of active galactic nuclei: the mean reddening of NGC~5548 and the size of the accretion disc}
\author[Gaskell et al.]{C. Martin Gaskell$^{1,}$\thanks{E-mail:
mgaskell@ucsc.edu}, Frances C. Anderson$^{1,2}$, Sufia Á. Birmingham$^{1,3}$, \and  Samhita Ghosh$^{1,4}$ \\
\\$^1$Department of Astronomy and Astrophysics, University of California, Santa Cruz, CA 95064
\\$^2$Department of Engineering, Harvey Mudd College,
Claremont, CA 91711
\\$^3$Department of Astrophysical Sciences
Princeton University, Princeton, NJ 08544
\\$^4$Department of Mathematics, University of California Berkeley, Berkeley, CA 94720-3840}

\begin{document}

\date{2022 August 24}

\pagerange{\pageref{firstpage}--\pageref{lastpage}} \pubyear{2021}

\maketitle

\label{firstpage}

\begin{abstract}
\\
We use seven different methods to estimate broad-line and continuum reddenings of NGC~5548.  We investigate two possible reddening curves considered for active galactic nuclei (AGNs): the mean AGN reddening curve of Gaskell \& Benker (2007) which is relatively flat in the ultraviolet, and a curve that rises strongly into the ultraviolet like a Small Magellanic Cloud (SMC) reddening curve.  We also consider a standard Milky Way curve.  Regardless of the curve adopted, we find a total reddening $\sim 14$ times greater than the small amount of reddening due to dust in the solar neighbourhood. The UV-to-optical ratios rule out a steep SMC-like reddening curve for NGC~5548.  The Milky Way and Gaskell \& Benker curves give a mean reddening of $E(B-V) = 0.25 \pm 0.02$. The four non-hydrogen-line reddening indicators imply that the intrinsic hydrogen line ratios are consistent with Baker-Menzel case B values. The unreddened optical to UV spectral energy distribution is consistent with the predicted distribution for an externally-illuminated accretion disc.  The reddening we derive for NGC 5548 is typical of previous estimates for type-1 AGNs.  Neglecting  internal extinction leads to an underestimate of the luminosity at 1200 \AA\ by a factor of seven.  The size scale of the accretion disc has therefore been underestimated by a factor of $\thickapprox 2.6$.  This is similar to the accretion disc size discrepancy found in the 2013 {\it AGNSTORM} campaign and thus supports the proposal by Gaskell (2017) that the accretion disc size discrepancy is primarily due to the neglect of reddening.
\end{abstract} 

\begin{keywords}
galaxies: active --- galaxies: nuclei --- quasars, emission lines --- galaxies: ISM --- dust, extinction --- galaxies: individual: NGC 5548
\end{keywords}

\section{Introduction}

Because it affects almost every aspect of our understanding of active galactic nuclei (AGNs), knowing the total reddening of an AGN is crucial.  The majority of studies of the spectral energy distributions (SEDs) and broad-line intensities of type-1 thermal AGNs assume that reddening other than by dust in our solar neighbourhood (typically $E(B-V) < 0.05$ for high Galactic latitude directions) is negligible.  On the other hand, other studies have pointed to substantial reddenings in some AGNs ($E(B-V) \sim 0.2$ and higher -- see \citealt{Gaskell17} for a review).  There are two main reasons why many researchers have assumed that internal reddening is low. The first is that there is only a small amount of additional reddening ($E(B-V) \sim 0.03 - 0.05$) associated with the  high-velocity outflows seen in broad absorption line quasars (BALQSOs), and the dust associated with the outflows has a reddening curve that rises strongly going into the UV like the reddening curve of the Small Magellanic Cloud (SMC) bar (see Figure 5 of \citealt{Gaskell+16}).  If $E(B-V) > 0.1$, and if the SMC-like reddening curve of BALQSOs is applicable to AGNs in general, the SED of AGNs should fall rapidly going into the far UV.  This is seen only rarely.  The second reason for thinking that the reddening of AGNs is low is that it has long been thought that in the broad-line region (BLR) of AGNs, Lyman $\upalpha$ is suppressed and the H$\upalpha$/H$\upbeta$ ratio increased by collisional and radiative transfer effects (e.g., \citealt{Ferland+Shields85}; for a review see  \citealt{Gaskell17}). If this is the case, observed Ly$\upalpha$/H$\upbeta$ and H$\upalpha$/H$\upbeta$ ratios imply little or no reddening.

To get around the potential problem with the intrinsic ratios of the BLR hydrogen lines we present here a study of seven different reddening indicators in spectra of NGC~5548, the best-studied, radio-quiet thermal AGN. Four of the reddening indicators do not involve hydrogen lines.  Over the past four decades the AGN NGC~5548 has been the subject of many large-scale reverberation mapping campaigns from the X-ray region to the IR. In analyzing recent results, \citet{Edelson+15}, \citet{Mehdipour+15}, \citet{Fausnaugh+16}, \citet{Kriss+19}, and others follow the widespread practice of assuming that there is negligible internal reddening along our line of sight and they only correct for reddening in the Milky Way.  The latter is very small.  \citet{Schlegel+98} get $E(B-V) = 0.020$ for NGC~5548, a value close to the earlier estimate of $E(B-V) = 0.017$ by \citet{Schlafly+Finkbeiner11}.  

In this paper we address five key issues: (1) are the different independent reddening indicators consistent? (2) are the hydrogen line ratios consistent with Case B ratio?  (3) what is the appropriate reddening curve for NGC~5548? (4) what is the average reddening of NGC~5548? and (5) can reddening solve the ``accretion-disc-size discrepancy" for NGC~5548? We also briefly consider other consequences of the reddening we find.  

The organization of the paper is as follows.  In Section 2 we discuss the general problem of estimating AGN reddening AGNs and consider three possible reddening curves.  In Section 3 we review and discuss four reddening indicators that do not involve hydrogen lines.  We give expected theoretical unreddened ratios and discuss the limitations of each indicator. In Section 4 we discuss the use of three hydrogen line ratios as reddening indicators.  In section 5 we use the seven methods to derive colour excesses, $E(\lambda_1 - \lambda_2)$, between pairs of wavelengths, $\uplambda_1$ and $\uplambda_2$, for each reddening indicator.  In Section 6 we compare the colour excesses $E(B-V)$ implied by the seven methods using the three reddening curves we consider.  In Sections 7 and 8 
we discuss the implications of the reddening we find for the UV luminosity and the size of its accretion disc respectively. In Section 9 we briefly discuss various consequences of our results and we end with some suggestions for optimizing future observing campaigns to improve reddening estimates.  Our main conclusions are summarized in Section 10.

\section{Estimating reddening of AGNs}

We follow the standard method of estimating reddenings.  We take observed ratios of the fluxes, $[F(\uplambda_1)/F(\uplambda_2)]_{\mathrm{obs}}$, at two different wavelengths, $\uplambda_1$ and $\uplambda_2$, and then make two assumptions.  The first is to assume an intrinsic (i.e., unreddened) flux ratio, $[F(\uplambda_1)/F(\uplambda_2)]_{\mathrm{int}}$. If $\uplambda_1 < \uplambda_2$, the colour excess between the two wavelengths is then
\begin{equation}
~~~~~~~~~~~E(\uplambda_1 -\uplambda_2) = -2.5 \log  \frac{[F(\uplambda_1)/F(\uplambda_2)]_{\mathrm{obs}}}{ [F(\uplambda_1)/F(\uplambda_2)]_{\mathrm{int}}} .
\end{equation}
Our second assumption is to adopt a reddening curve to get standard reddenings in terms of $E(B-V)$, the colour excess between the Johnson $B$ and $V$ photometric bands. With one exception, the reddening indicators we consider are between other pairs of wavelengths.  To convert a colour excess, $E(\lambda_1-\lambda_2)$, to an equivalent $E(B-V)$ one needs a reddening  curve, $A(\uplambda)/E(B-V)$, where $A(\uplambda)$ is the relative extinction in magnitudes at wavelength $\uplambda$.  From the adopted reddening curve one can determine

\begin{equation}
~~~~~~~~~~~\frac{E(\uplambda_1-\uplambda_2)}{E(B-V)} = \left[ \frac{A(\uplambda_1)}{E(B-V)}\right] - \left[\frac{A({\uplambda_2})}{E(B-V)}\right], 
\end{equation}
where the quantities in square brackets on the right are the normalized attenuations for the relevant wavelengths.  These are read off from the normalized reddening curve being used.  One can thus convert $E(\uplambda_1,\uplambda_2)$ into $E(B-V)$.

\begin{table*}
\caption{Reddening indicators and reddening curves}
\label{symbols}
\begin{tabular}{lcccccc}
\hline																	
Method	&	$\uplambda_{\rmn{short}}$	&	$\uplambda_{\rmn{long}}$	& Intrinsic &	$\frac{E(\uplambda_1-\uplambda_2)}{E(B-V)}$	& 	$\frac{E(\uplambda_1-\uplambda_2)}{E(B-V)}$	& 	$\frac{E(\uplambda_1-\uplambda_2)}{E(B-V)}$ \\
& & & ratio &  Milky Way & Gaskell \& Benker & SMC \\
\hline																	
Ly$\upalpha$/H$\upbeta$	&	$\uplambda$1216	&	$\uplambda$4861	&	35	&	6.75	& 4.90 & 16.7  \\
\ion{O}{i}	&	$\uplambda$1304	&	$\uplambda$8446	&	6.5	&	7.64 & 6.75 & 17.2  	\\
$F_{\mathrm{uv}}/F_{\mathrm{opt}}$	&	$\uplambda$1337	&	$\uplambda$5100	&	1.5	&	5.68 & 5.03 & 14.8  		\\
\ion{He}{ii}	&	$\uplambda$1640	&	$\uplambda$4686	&	8.5	 &	3.99 & 4.26 & 9.83  		\\
$F_B/F_V$	&	$\uplambda$4450	&	$\uplambda$5510 &	1.18	& 1.00 & 1.00 & 1.00  	\\
H$\upalpha$/H$\upbeta$	&	$\uplambda$4861	&	$\uplambda$6563	&	2.8	&	1.23	& 1.23 & 1.20  	\\
Pa$\upbeta$/H$\upbeta$	&	$\uplambda$4861	&	1.28 $\mu$m	&	0.14	&	2.81 & 2.76 & 2.71  	\\
\hline																				
\end{tabular}
\end{table*}

\subsection{Reddening curves}

The shape of a reddening curve depends on the composition and size distribution of the grains and on the geometry of the dust region. The choice of curve is relatively unimportant in the optical and IR because curves are fairly similar, but the choice strongly effects the value of $E(B-V)$ derived from UV-to-optical reddening indicators because there are large differences in reddening curves at short wavelengths.  The steeper the reddening curve is in the UV, the lower $E(B-V)$ is for a given $E(UV-optical)$. 

\citet{Wampler68} presciently noted that ``if the dust causing the reddening [of AGNs] is produced in the nuclei of the Seyfert galaxies, there is no reason to believe that it would have the same reddening properties as dust produced in the spiral arms of our Galaxy." \citet{Gaskell+04} discovered from comparing composite spectra of radio-loud AGNs at different orientations that the reddening curve for these AGNs was substantially flatter in the UV than a typical Milky Way curve in the UV.  A subsequent independent analysis by \citet{Czerny+04} of large sample of predominantly radio-quiet SDSS AGNs also gave a reddening curve that was flatter than a typical Milky Way curve in the UV.  In contrast to these results, \citet{Gaskell+16} have shown that a reddening curve which steeply rises to the far UV, like the reddening curve of the bar of the SMC bar, is an excellent fit to the reddening curve of dust associated with high-velocity outflows from AGNS (the broad absorption lines). \citet{Gaskell+Benker07} constructed reddening curves for 14 AGNs with near-simultaneous observations from the far UV to long optical wavelengths.  These show some variety in the reddening curves, but only one of the 14 AGNs had a steep SMC-like reddening curve.\footnote{Sample selection has an important effect.  Samples of AGNs selected by having red UV colors will naturally produce more SMC-like reddening curves.  This is the case, for example, for the red AGNs considered by \citet{Richards+03}.}  

In this paper we will consider three reddening curves.  The first is a typical Milky Way reddening curve, as parametrized by \citet{Weingartner+Draine01}.  A Milky Way curve has long been ruled out for the majority of AGNs because they do not show the strong $\uplambda$2175 dust feature, so we are only including the Milky Way curve for comparison.  We then consider two reddening curves proposed for AGNs: the mean AGN reddening curve given by \citet{Gaskell+Benker07}, and an SMC-like curve as parametrized by \citet{Weingartner+Draine01}. \citet{Gaskell+Benker07} only gave their mean AGN curve for wavelengths from Lyman $\upalpha$ to H$\upalpha$, but we consider Paschen $\upbeta$ here as well.  Since the SMC and Milky Way curves are similar from H$\upalpha$ to Pa$\upbeta$, we have extended the Gaskell \& Benker curve to longer wavelengths by taking the average of the Milky Way and SMC curves.
Table 1 summarizes our adopted factors for the three reddening curves for converting $E(\uplambda_1-\uplambda_2)$ to $E(B-V)$ for the pairs of wavelengths we consider (i.e., the right hand side of Eq. 2)

\section{Non-hydrogenic reddening indicators}

Because of the long-standing controversy over hydrogen line ratios (see Section 4), we first consider four reddening indicators that do {\em not} involve hydrogen lines.  We discuss some of the limitations of each indicator and questions that have been raised.

\subsection{\ion{He}{II} lines}

\citet{Shuder+MacAlpine79} pointed out that, in principle, one of the best reddening indicators for the BLR is the \ion{He}{ii} $\uplambda$1640/$\uplambda$4686 line ratio. Because the abundance of helium by number is a tenth that of hydrogen, and because the lines arise from very high energy levels in gas of high ionization, the \ion{He}{ii} $\uplambda$1640/$\uplambda$4686 ratio is expected to be much less sensitive to the collisional-excitation and optical-depth effects which had been suggested to potentially drastically change hydrogen line ratios \citep{MacAlpine81}.  Photoionization models of the high-ionization BLR \citep{MacAlpine81,Bottorff+02} confirm that the \ion{He}{ii} $\uplambda$1640/$\uplambda$4686 ratio is close to a Baker-Menzel case B value. We adopt a theoretical $\uplambda$1640/$\uplambda$4686 intensity ratio of $8.5 \pm 1$ for the high-ionization BLR (see Figures 2e and 3 of \citealt{Bottorff+02}).  The \ion{He}{ii} $\uplambda$1640/$\uplambda$4686 ratio is not expected to vary much with physical conditions because the lines arise from high levels.

Unfortunately, both \ion{He}{ii} $\uplambda$1640 and $\uplambda$4686 are relatively weak, blended lines.  This has deterred their use as a reddening indicator.  \ion{He}{ii} $\uplambda$1640 is blended with \ion{O}{iii}] $\uplambda$1663 and the red wing of \ion{C}{iv} $\uplambda$1549; \ion{He}{ii} $\uplambda$4686 is blended with broad \ion{Fe}{ii} $\uplambda$4570 emission that can often be strong enough to make \ion{He}{ii} $\uplambda$4686 unrecognizable.  \ion{He}{ii} $\uplambda$4686 can also blend with the blue wing of the strong H$\upbeta$ line.  Stellar absorption lines from the host galaxy are a further complication for \ion{He}{ii} $\uplambda$4686.  A problem with measuring \ion{He}{ii} lines (especially $\uplambda$1640) is that higher-ionization lines are progressively broader than lower-ionization lines \citep{Shuder82,Mathews+Wampler85,Krolik+91} because the BLR is strongly radially stratified by ionization (see \citealt{Gaskell09}). Because of this, \ion{He}{ii} cannot be de-blended by assuming that a \ion{He}{ii} line has the same profile as the contaminating lines.

\ion{He}{ii} arises in the innermost BLR close to the innermost part of the accretion disc. Because it takes 54 eV to doubly ionized helium, most of the ionizing photons needed to produce \ion{He}{ii} come from close to the Wien cutoff of the spectrum of the accretion disc. This emission from the innermost accretion disc is the most variable part of an AGN's spectrum. \ion{He}{ii} comes from ten or more times closer to the black hole than other BLR lines studied (see \citealt{Gaskell09} for a review of the structure of the BLR).  Because of the combination of the small size of the He$^+$ emitting region and the rapid, high variability of the relevant ionizing continuum, \ion{He}{ii} emission is highly variable. For NGC~5548, \ion{He}{ii} variations lag continuum variations by less than a day \citep{Korista+95} rather than the many days or weeks for other broad lines.  The \ion{He}{ii} lines can only be used as a reddening indicator if the \ion{He}{ii} $\uplambda$1640 and $\uplambda$4686 observations are near simultaneous.  We will propose in Section 5.1 below that the strong variability of \ion{He}{ii} can be exploited to overcome the line-blending problem and estimate the $\uplambda$1640/$\uplambda$4686 ratio.

\subsection{O\,I lines}

\citet{Netzer+Davidson79} proposed that the \ion{O}{i} $\uplambda$1304/$\uplambda$8446 ratio is another useful reddening indicator since almost every $\uplambda$8446 transition will be followed directly by a $\uplambda$1304 transition.  The number of $\uplambda$1304 photons is the same as the number of $\uplambda$8446 photons but the intensity of the  $\uplambda$8446 line is stronger because a $\uplambda$1304 photon has more energy than a $\uplambda$8446 one.  This thus gives an \ion{O}{i} $\uplambda$1304/$\uplambda$8446 intensity ratio of 8446/1304 = 6.48. \citet{Kwan+Krolik81} and \citet{Grandi83} noted minor effects that might possibly change the $\uplambda$1304/$\uplambda$8446 ratio (see \citealt{Grandi83}, section IId).  Various factors which might affect the \ion{O}{i} $\uplambda$1304/$\uplambda$8446 ratio were also discussed by \citet{Rodriguez-Ardila+02} and modelled by \citet{Matsuoka+07}.  The latter concluded that the ratio was close to 6.48 for high-column-density clouds with densities and ionization parameters similar to those expected for the BLR.  The situation in real BLRs will be even better because observations imply that BLR emission is spread out and highly radially stratified by ionization \citep{Gaskell+07,Gaskell09}, rather than coming from high-column-density clouds where each cloud emits all lines.  Furthermore, photons will escape from the sides of clouds (see Figure 3 of \citealt{Gaskell17}).  We therefore expect the unreddened theoretical \ion{O}{i} $\uplambda$1304/$\uplambda$8446 ratio to be close to 6.48.

\citet{Baldwin+96} suggested that much of what has been assumed to be the \ion{O}{i} $\uplambda$1304 quartet could be the \ion{Si}{ii} $\uplambda$1307 doublet and
that the latter could dominate in many objects.  \citet{Matsuoka+07} propose that typically \ion{Si}{ii} makes up 50\% of the blend.  If this is the case, assuming that the observed $\uplambda$1304/$\uplambda$8446 ratio is just giving the ratio of the \ion{O}{i} lines will result in a major underestimate of the reddening.   However, such strong \ion{Si}{ii} $\uplambda$1307 is a problem for emission line modelling because photoionization models cannot produce strong \ion{Si}{ii} $\uplambda$1307. (\citealt{Baldwin+96} refer this this as ``the \ion{Si}{ii} disaster'' -- see their Appendix C). We will nevertheless investigate the question of possible \ion{Si}{ii} contamination in Section 5.2.

\subsection{Reddening from the shape of the variable optical flux (Cho{\l}oniewski method)}

\citet{Choloniewski81} made the important discovery that if the apparent brightnesses of an AGN in different passbands -- for example, the fluxes $F_B$ and $F_V$ in the Johnson $B$ and $V$ bands -- are plotted against each other as {\em fluxes} rather than as magnitudes, the curvature seen when plotting in magnitudes disappears. He interpreted the resulting straight lines in flux--flux plots\footnote{Note that although the standard procedure has been to determine gradients from flux--flux plots, the gradient can also be obtained from root-mean-square spectra.} as the addition of a non-varying component (mostly starlight of the host galaxy) and a varying AGN component of fixed spectral shape.  He further suggested that differences in the slope in flux-variability plots were a consequence of differing reddening in each AGN.  Careful observations and analysis by \citet{Winkler+92}, \citet{Winkler97} and \citet{Sakata+10} have shown that the loci in optical flux-variability plots are indeed straight lines.  \citet{Winkler+92}, \citet{Cackett+07} and \citet{Heard+Gaskell22} also show that there is reasonable agreement with other reddening indicators, thus supporting Cho{\l}oniewski's conjecture.

\subsubsection{The unreddened optical flux}

To estimate $E(B-V)$ via the Cho{\l}oniewski method one needs to know the spectral shape of the variable component of an unreddened AGN. The optical-to-UV emission of a thermal AGN\footnote{See \citet{Antonucci12} for discussion of the division of AGNs into thermal (high accretion rate) and non-thermal (low accretion rate) AGNs.} is dominated by thermal emission from the accretion disc.  The outer parts of the disc are externally illuminated by emission just above the disc plane closer to the centre of the disc.  This produces a spectrum with a power-law dependence of the flux per unit frequency on the frequency: $F_{\nu} \propto \nu^{+1/3}$ \citep{Friedjung85}\footnote{Coincidentally and confusingly, this is also the better-known theoretical spectrum produced by the outer parts of the standard internally-heated accretion disc of \citet{Lynden-Bell69}.}.  \citet{Heard+Gaskell22} show that the distribution of spectral indices of the variable component of the optical-to-near-UV spectra in over 4000 SDSS AGNs is consistent with the unreddened continuum having the expected $\upnu^{+1/3}$ spectrum after allowance for BLR emission.  The intrinsic ratio of the unreddened fluxes per unit frequency between two wavelengths of the disc component alone is thus expected to be

\begin{equation}
~~~~~~~~~~~~~~~~~~~~\left[ \frac{F(\uplambda_1)}{F(\uplambda_2)}\right]_{\mathrm{int}} = \left[\frac{\uplambda_2}{\uplambda_1}\right]^{+1/3} .
\end{equation}

Strictly speaking, the theoretical ratio given by Eq.~3 only gives the slope in a flux-variability diagram when {\em monochromatic} fluxes are plotted, such as fluxes obtained from spectra after allowance for line emission.  If the method is being used with fluxes found from broad-band filter measurements (as is frequently the case), contamination by variable line emission and bound-free emission associated with the BLR causes the unreddened SED to deviate somewhat from the expected theoretical spectrum for a pure externally-illuminated accretion disc (see \citealt{Heard+Gaskell22}).   Depending on where variable strong emission lines and bound-free continua fall with respect to filter passbands, the unreddened slope will be more or less than the slope given by Eq.~3. For low-redshift thermal AGNs, only the $V$ filter does not include any strong contribution from a broad line.  The $R$ filter, for example, is centred on the very strong, broad H$\upalpha$ line.  This will make the unreddened $F_V$ vs.\@ $F_R$ gradient much less (i.e., redder) than the value given by Eq.\@3.  The $R$ filter, and especially the $I$ filter, also include emission from the hot nuclear dust (see \citealt{Gaskell07} and \citealt{Sakata+10}). At shorter wavelengths, the $U$ filter has a strong contribution from the BLR emission making up the ``small blue bump" (see Figure 4 of \citealt{Heard+Gaskell22}).  After the $V$ band, the filter with the least broad-line emission for low-redshift AGNs is the $B$ filter but this includes the higher-order Balmer lines blending into the Balmer continuum emission of the long wavelength side of the small blue bump. The unreddened $F_B$ vs.\@ $F_V$ flux gradient will thus be greater than the value of 1.07 predicted by Eq.\@ 3.  Comparing an observed $F_B$ vs.\@ $F_V$ gradient with 1.07 therefore gives only a {\em lower limit} to $E(B-V)$.  The effect of the broad-line contamination on the $B$ band is illustrated by the distribution of $F_B$ vs.\@ $F_V$ gradients found by \citet{Winkler97} for a large sample of low-redshift AGNs.  Fully 40\% of his $F_B$ vs.\@ $F_V$ gradients are greater than the 1.07 predicted by Eq.\@ 3. 

One can get an empirical estimate of the unreddened $F_B$ vs.\@ $F_V$ gradient of a typical low-redshift AGN by considering the bluest variable AGN continua.  One cannot simply take the bluest known gradient because extrema are influenced by observational errors.  An accurate estimation of the bluest $(B-V)$ depends on knowing the errors in the gradient estimates, something beyond the scope of this study.  However, the 90th percentile of the $F_B$ vs.\@ $F_V$ gradients of the large \citet{Winkler97} low-redshift sample is 1.18.  We will therefore adopt this as the unreddened $F_B$ vs.\@ $F_V$ gradient.  The 80th percentile of the distribution gives 1.12, which suggests the uncertainty in the assumed unreddened slope is of the order of $\pm 0.06$.  Assuming an intrinsic slope of 1.18 compared with 1.07 raises an estimate of $E(B-V$ by $\sim 0.11$ magnitudes.  \citet{Cackett+07} noted that the reddenings they got from the Cho{\l}oniewski method were systematically lower that the reddenings implied by the Balmer decrements.  Increasing their reddenings from flux-variability gradients by 0.11 magnitudes, as suggested here, removes this systematic difference.

In addition to using an incorrect unreddened flux gradient, another factor that will give an incorrect reddening is having a scale-factor error in one of both of the fluxes.  Whilst the Cho{\l}oniewski method automatically removes constant additive components, scale factor errors will give an incorrect slope.  When converting broad-band magnitudes to fluxes it is therefore important to use the correct conversion factor for each of the photometric filters used to make the observations.

\subsection{UV-to-optical continuum flux ratio}

The ratio of the UV to optical flux is strongly influenced by reddening.  Whilst it is well established that there is little if any variability in the intrinsic SED in the {\em optical} as an AGN varies (see previous section), this is not necessarily expected to be the case for the shortest wavelengths. This is because of two effects.  Firstly, as one goes to shorter wavelengths, one eventually approaches the Wien cutoff of emission from the  innermost regions of the disc. This causes a flattening of the SED towards shorter wavelengths and a departure from a $F_{\nu} \propto \nu^{+1/3}$ power law at short wavelengths.  The continuum variability of high-luminosity AGNs \citep{Heard+Gaskell22} suggests that there is indeed happening at least for high-luminosity AGNs (see Figure 2 of \citealt{Heard+Gaskell22}).  The second effect is that, since the heating of the surface of the accretion disc comes from external illumination, geometric effects alone flatten the rise in temperature with decreasing radius as one approaches the disc centre.  This too causes a flattening of the SED at very short wavelengths.  The extent to which this flattening varies from object to object as a function of parameters such as black hole mass, luminosity and Eddington ratio needs further investigation but the study by \citet{Heard+Gaskell22} of continuum variability going out to $\sim 1200$\AA\ indicates that, after allowance for the ``small blue bump" and other emission associated with the BLR, the Cho{\l}oniewski method can be used down to at least 2200\AA.  

Comparing the variability of H$\upbeta$ with predictions from variability of the continuum measured at 5100 \AA\ shows that they are frequent deviations from the predictions \citep{Gaskell+21}.  These ``anomalies" are on a timescale of $\sim 40 \pm 20$ days or so (see Figure 4 of \citealt{Gaskell+21}) and are independent of the continuum level. Since H$\upbeta$ is responding to variations in the ionizing flux with wavelengths shorter than 912 \AA , the anomalies in the H$\upbeta$ response must correspond to changes in the UV to optical continuum shape.  Because of these it is important not only to try to determine the UV/optical flux ratio from near-simultaneous observations, but also to to average over variability.

\begin{figure} 
	\label{fig:LyaHbContinuum}
	\begin{center}
		\includegraphics[width = 1.10\linewidth]{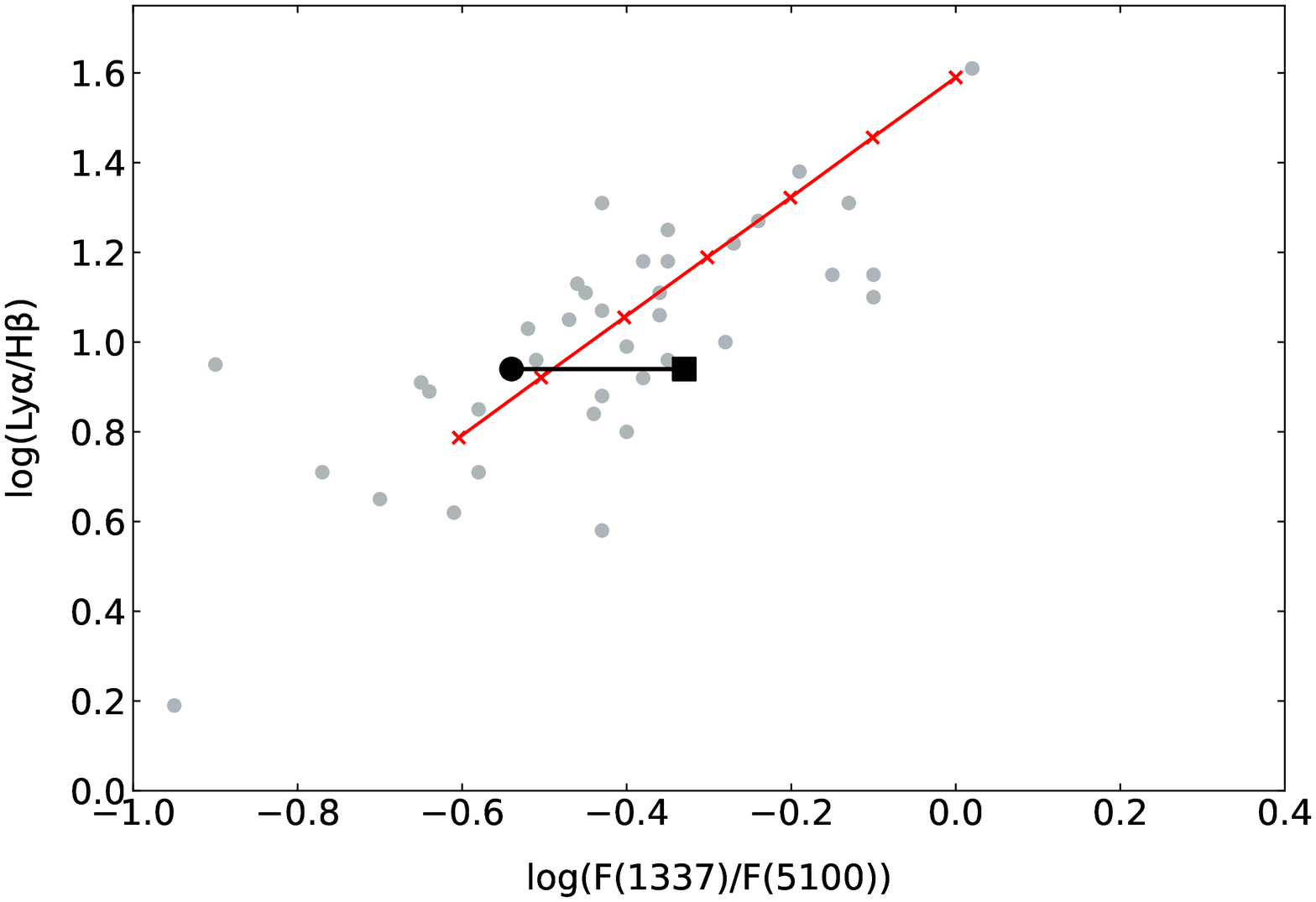}
        \caption{Ly$\upalpha$/H$\upbeta$ ratios versus the ratio of UV to optical continuum fluxes.  The gray dots are the AGN measurements reported by \citet{Netzer+95} and \citet{Bechtold+97}. The contribution of host galaxy starlight to the optical fluxes has not been allowed for.  The filled black circle shows where NGC~5548 falls without removal of the contribution of the host galaxy starlight to the optical flux. The black square is NGC~5548 with the host galaxy starlight removed. The red line is the reddening vector for the \citet{Gaskell+Benker07} curve (see text) with crosses separated by 0.05 magnitudes in $E(B-V)$ and starting in the upper right at $E(B-V) = 0.00$ in the upper right. (Figure adapted from Figure 2 of \citealt{Gaskell17}.)} 
	\end{center}
\end{figure}

Despite these expected problems, studies by \citet{Netzer+95} and \citet{Bechtold+97} showed empirically that the $F_{\uplambda1216}/F_{\uplambda4861}$ continuum flux ratio is correlated with the Ly$\upalpha$/H$\upbeta$ line flux ratio.  Figure 1 shows the combined data sets. Such a correlation is a natural consequence of reddening, but there is no ready explanation of why the intrinsic Ly$\upalpha$/H$\upbeta$ line ratio might depend on the optical-to-UV spectral slope. This therefore implies that the UV/optical flux ratio can be used at least as an approximate reddening indicator.  We assume that the unreddened Ly$\upalpha$/H$\upbeta$ ratio is close to Case B (see next section) and plot a reddening vector starting at this value.  We have adjusted the vector in the continuum flux direction so that it passes through the median of the other AGNs.  If we do this, the continuum flux ratio is only slightly less than the value of $F_{\uplambda1216}/F_{\uplambda4861} = (4861/1216)^{1/3}$ predicted by the extrapolation of the $F_{\nu} \propto \nu^{+1/3}$ power-law expected at longer wavelengths.   A further cause of uncertainty is that the optical fluxes plotted in Figure 1 are not corrected for the unknown contributions of host galaxy starlight.  Correction for these will lower the optical fluxes somewhat and hence increase the UV/optical flux ratio to bring the ratio for the bluest AGNs closer to $(4861/1216)^{1/3}$. Because of the uncertainty of the host galaxy starlight contribution, an $F_{\uplambda1216}/F_{\uplambda4861}$ ratio of 1.2 is not inconsistent with the prediction of 1.5 one gets if the $F_{\nu} \propto \nu^{+1/3}$ spectrum of an externally-illuminated accretion disc is extrapolated down to $\uplambda$1300. We discuss the size of the host galaxy starlight correction and the location of the ratios for NGC~5548 in Section 5.4 below.

\section{Hydrogen line ratios}

We now consider three hydrogen line ratios.  The intrinsic hydrogen line ratios have been controversial since the first spectrophotometry of AGNs in the mid-1960s. This controversy, which has been entwined with the AGN reddening question, is reviewed in \citet{Gaskell17}.  In the 1960s, low Ly$\upalpha$/H$\upbeta$ and high H$\alpha$/H$\upbeta$ ratios were found which differed from the ratios expected from Case B values of recombination theory \citep{Wampler67,Wampler68}.  \citet{Wampler67} noted that reddening of $E(B-V) \sim 0.2$ could explain the Balmer decrement of Ton 1542.  He also estimated the Ly$\alpha$/H$\alpha$ ratio as $\approx 2.7$ by comparing high- and low-redshift AGNs \citep{Wampler68}.  This was much less than the Case B prediction.  Although reddening offered a solution to the differences from Case B, theoretical calculations by many research groups showed that some of the observational results {\em could} be explained by radiative transfer and collisional effects instead, but it was difficult to obtain a coherent picture.  \citet{Gaskell17} presented evidence that, after de-reddening, velocity-integrated broad-line hydrogen line ratios in AGNs were consistent with Case B recombination values.  He pointed out that optically-thick, one-dimensional-slab radiative transfer models that produce values differing from Case B were unlikely to be relevant to real AGNs because the line radiation would escape from clouds sideways.

\subsection{H$\upalpha$  / H$\upbeta$}

For the {\em narrow}-line region the Balmer decrement, H$\upalpha$/H$\upbeta$, is an easy-to-use reddening indicator.  The intrinsic ratio is expected to be close to a Menzel \& Baker Case B value of 2.7--3.1 (see Table 4.4 of \citealt{Osterbrock+Ferland06}). Comparison with other NLR reddening indicators \citep{Gaskell82,Gaskell84,Wysota+Gaskell88} supports this. However, as noted, for the {\em broad}-line region there have long been doubts about whether Balmer-line ratios are compatible with simple reddening of case B values.  There are two firm observational indications that the intrinsic ratio {\em does} depend somewhat on physical conditions.  Firstly, the Balmer decrement at a given time becomes flatter with increasing gas velocity \citep{Shuder82,Crenshaw86}.  Secondly, reverberation mapping shows that the variability-weighted radius of the higher-order Balmer lines is progressively somewhat smaller as one goes up the Balmer series (e.g., \citealt{Kaspi+00}).  These results point to the Balmer decrement getting flatter with increasing density and temperature as is expected closer to the centre of the AGN.  This is in agreement with recombination theory which predicts that the Balmer decrement gets flatter with increasing temperature and increasing density.  This can be seen in Table 4.4 of \citealt{Osterbrock+Ferland06}.  Figure 4 of \citet{Gaskell17} shows the effect for higher densities.  Despite these expected and observed effects, the {\em velocity-integrated} line ratios of the bluest, and presumably least-reddened, AGNs nevertheless imply an intrinsic H$\upalpha$/H$\upbeta$ ratio of $\thickapprox 2.8$ for the BLR as a whole \citep{Gaskell17}.  This is close to a Case B value.  We will therefore adopt H$\alpha$/H$\upbeta$ = 2.8 for the velocity-integrated broad lines of NGC~5548.

\subsection{Lyman $\upalpha$ / H$\upbeta$}

From the correlation between Ly$\upalpha$/H$\upbeta$ and the UV/optical flux ratio (see Figure 1 and Section 3.4 above) \citet{Gaskell17} argued empirically that the velocity-integrated Ly$\upalpha$/H$\upbeta$ flux ratio is close to a Case B value of $\thickapprox 35$.  \citet{Zheng92} found that the Ly$\upalpha$/H$\upbeta$ ratio is higher in the high-velocity wings of the lines. \citet{Gaskell17} proposed that the Lyman $\upalpha$ enhancement at high velocities is due to continuum ﬂuorescence in the Lyman lines (so-called ``Case C" conditions -- see \citealt{Baker+38}, \citealt{Chamberlain53}, and \citealt{Ferland99}).  Because of this enhancement of Lyman $\upalpha$ in the high-velocity wings, the velocity-integrated intrinsic Ly$\upalpha$/H$\upbeta$ ratio could be somewhat greater than a Case B value, but for NGC~5548 we will assume a Case B Ly$\upalpha$/H$\upbeta$ ratio of 35.

\subsection{Paschen $\upbeta$ / H$\upbeta$}

Pa$\upalpha$ and Pa$\upbeta$ are in the near IR and thus comparing them with the Balmer lines samples a different part of the reddening curve than UV to optical ratios. Because Pa$\upalpha$ and Pa$\upbeta$ arise from higher levels of hydrogen than Lyman $\upalpha$ and the H$\alpha$ and have lower transition probabilities, there is less danger of the Paschen lines being affected by collisional and optical-depth effects than H$\upalpha$.   The intrinsic Pa$\upbeta$/H$\upbeta$ ratio is expected to be $\thickapprox 0.14$.

\section{Estimating the reddening of NGC~5548}

We now use these seven reddening indicators to estimate the reddening of NGC~5548.

\subsection{He II $\uplambda$1640/$\uplambda$4686}

As noted in Section 3.1, both the $\uplambda$1640 and $\uplambda$4686 lines of \ion{He}{ii} suffer from severe blending with other lines.  However, one can use the strong, rapid variability of \ion{He}{ii} to reduce the effects of contamination of the \ion{He}{ii} $\uplambda$1640 and $\uplambda$4686 lines by other spectral features.  Because the ionization potentials of oxygen are only slightly higher the those of carbon for the same stages of ionization, \ion{O}{iii}] $\uplambda$1663, which blends with \ion{He}{ii} $\uplambda$1640, is produced in approximately the same region as \ion{C}{iii}] $\uplambda$1909.  Photoionization models show that \ion{C}{iii}] $\uplambda$1909 is produced at column densities at order of magnitude higher than \ion{He}{ii}.  The self-shielding model of \citet{Gaskell+07} hence predicts that \ion{O}{iii}] will be emitted at a radius an order of magnitude larger than \ion{He}{ii}.  This is strongly supported by reverberation-mapping.  For NGC~5548, \citet{Clavel+91} get a \ion{C}{iii}] $\uplambda$1909 lag of 34 days behind the continuum whilst \citet{Korista+95} get a \ion{He}{ii} lag of $< 2$ days. \ion{O}{iii}] $\uplambda$1663 therefore cannot follow rapid variability of the ionizing continuum the way \ion{He}{ii} $\uplambda$1640 can and does. This is true {\it a fortiori} for the optical \ion{Fe}{ii} emission blended with \ion{He}{ii} $\uplambda$4686 because \ion{Fe}{ii} arises in the lowest ionization part of the BLR with an effective radius twice that of H$\upbeta$ \citep{Gaskell+22}.  For NGC~5548, \ion{Fe}{ii} variability will lag the continuum variability by $\sim 40$ days.  Given this much slower response of contaminating lines in the UV and optical, we propose two methods that exploit the stronger variability of \ion{He}{ii} for estimating the contamination-free \ion{He}{ii} $\uplambda$1640/$\uplambda$4686 ratio.

\subsubsection{Difference spectra of He II lines}

A ratio of \ion{He}{ii} $\uplambda$1640/$\uplambda$4686 can be found by comparing {\em difference spectra} between high and low states of \ion{He}{ii} emission rather than from spectra themselves.  Because the lower-ionization lines blending with \ion{He}{ii} will not vary as much as \ion{He}{ii}, difference spectra between high and low states will mostly show the highly variable \ion{He}{ii} line. The variable component of the \ion{He}{ii} lines can thus readily be identified by subtracting a low state from a high state for which simultaneous optical and UV spectra are available.  During the 1993 {\it International AGN Watch} monitoring campaign \citep{Korista+95} simultaneous ($\pm 1$d) {\it HST} and optical spectra were obtained on five days.  Figure 2 shows continuum-subtracted UV and optical difference spectra for the regions around the \ion{He}{ii} lines for JD 2449128 and JD 2449099.  It can be see that this has removed all of the \ion{O}{iii}] and broad optical \ion{Fe}{ii} emission and that the difference profiles for the two \ion{He}{ii} lines agree well.  The observed ratio is 4.  For an intrinsic ratio of 8.5 this gives $E(\uplambda1640-\uplambda4686) = 0.82$ magnitudes

\begin{figure} 
	\label{fig:ContinuumSubtractedHeII}
	\begin{center}
		\includegraphics[width = 1\linewidth]{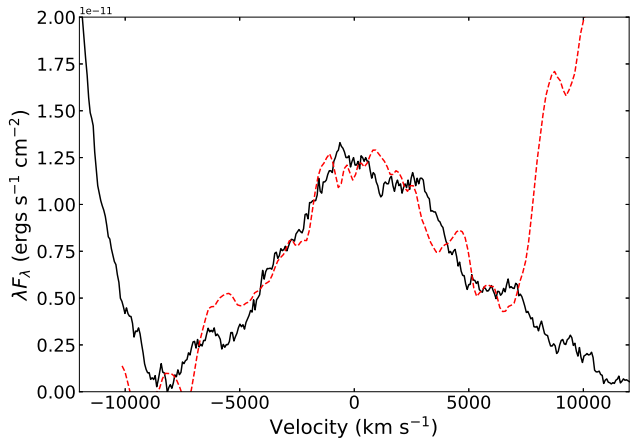}
        \caption{Continuum-subtracted UV and optical \ion{He}{ii} difference spectra between JD 2449128 and JD 2449099.  The thick solid line is \ion{He}{ii} $\uplambda$1640 and the dashed (red) line is \ion{He}{ii} $\uplambda$4686.  The optical spectrum has been multiplied by 4.0 to make the two \ion{He}{ii} profiles match.  The strong UV line bluewards of -9000 km~s$^{-1}$ is \ion{C}{iv} $\uplambda$1549 and the strong optical line redwards of +7000 km~s$^{-1}$ is H$\upbeta$.} 
	\end{center}
\end{figure}

\subsubsection{Line--line variability plots}

The other method we propose for exploiting variability to estimate the \ion{He}{ii} $\uplambda$1640/$\uplambda$4686 ratio is to look at the slopes of line flux plots of the different lines.  Because $\uplambda$1640 and $\uplambda$4686 are from the same ion, {\em their intensities must be directly proportional.}  The slope of the line in the flux-flux plot gives the mean line ratio. We show examples of \ion{He}{ii} line flux plots in Figures 3 and 4.  Just as non-variable host galaxy starlight in the Cho{\l}oniewski method does not affect the slope in a plot of varying continuum fluxes, so contaminating non-variable line emission does not affect the slope in a plot of varying line fluxes.  Contamination simply produces a constant offset. One caveat is necessary though: whilst constant or nearly constant contamination cancels out, systematic scale factor errors in measuring one or both of the lines will not.  A scale factor error in one or both fluxes will modify the slope and change the estimated of the line ratio.

\citet{Clavel+91} and \citet{Dietrich+93} give intensities of \ion{He}{ii} $\uplambda$1640 and $\uplambda$4686 respectively for the 1988-1989 {\it International AGN Watch} NGC~5548 campaign. The published measurements include contamination from other lines which have not been removed.  In Figure 3 we show \ion{He}{ii} $\uplambda$1640 and $\uplambda$4686 intensities observed within $\pm 2$ days.\footnote{\citet{Clavel+91} give fluxes measured with two different extraction routines.  We used the fluxes found from the SIPS extraction since they give better consistency between closely-spaced observations.}  The expected intrinsic unreddened \ion{He}{ii} $\uplambda$1640/\ion{He}{ii} $\uplambda$4686 ratio of 8.5 is indicated by the diagonal dashed line.  Since quantities plotted in Figure 3 and Figure 4 have errors in both axes, the fits are the ordinary least squares bisector (OLS bisector) advocated by \citet{Isobe+90}. For display purposes, to emphasize that what matters is the slope, we have slightly reduced the \ion{He}{ii} $\uplambda$4686 fluxes by $10 \times 10^{-15}$ ergs s$^{-1}$ cm$^{-2}$ to make the OLS-bisector line pass through the origin.  The offset is unimportant; what matters is the slope.  The slope of the OLS-bisector gives an observed mean $\uplambda$1640/$\uplambda$4686 intensity ratio of $3.29 \pm 0.12$.   Comparison with the expected intrinsic ratio of $8.5 \pm 1$ gives a mean $E(\uplambda1640-\uplambda4686)$ of $1.03 \pm 0.19$ magnitudes. The estimated error includes the statistical error in the OLS-bisector slope and the estimated uncertainty in the intrinsic ratio (see \citealt{Bottorff+02}).

\begin{figure} 
	\label{fig:HeIIIAWVectors}
	\begin{center}
		\includegraphics[width = 1\linewidth]{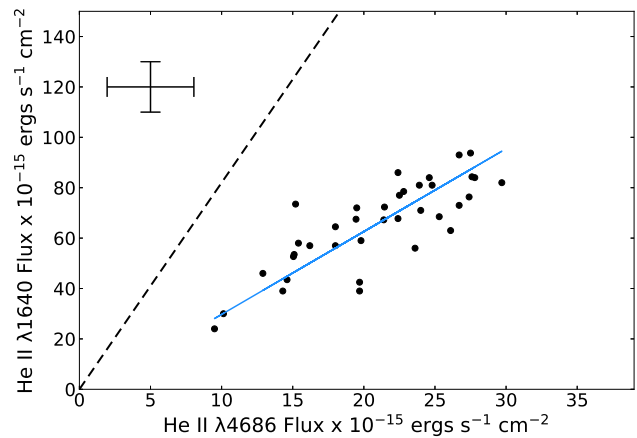}
        \caption{Near-simultaneous \ion{He}{ii} $\uplambda$1640 vs \ion{He}{ii} $\uplambda$4686 fluxes from the 1988-1989 {\it International AGN Watch} campaign. The $\uplambda$1640 measurements are the SIPS extracted fluxes from \citet{Clavel+91} and have not had \ion{O}{iii} $\uplambda$1663 emission removed. The \ion{He}{ii} $\uplambda$4686 has not had \ion{Fe}{ii} emission removed.  The dashed line is a theoretical \ion{He}{ii} $\uplambda$1640/$\uplambda$4686 ratio of 8.5 and the solid (blue) line is an OLS-bisector fit to the data.  The \ion{He}{ii} $\uplambda$4686 fluxes have been reduced by $10 \times 10^{-15}$ ergs s$^{-1}$ cm$^{-2}$ so that this line goes through the origin.}
	\end{center}
\end{figure}

Near-simultaneous \ion{He}{ii} $\uplambda$1640 and $\uplambda$4686 fluxes have also been reported from the 2013 {\it AGNSTORM} campaign by \citet{deRosa+15} and \citet{Pei+17} respectively.  The uncorrected values as reported are shown as the open circles in Figure 4.  Taken at face value these straddle the theoretical ratio and this could be taken as evidence for no reddening.  However, whilst \citet{Pei+17} subtracted out broad optical \ion{Fe}{ii} emission from \ion{He}{ii} $\uplambda$4686, \citet{deRosa+15} did not subtract out the substantial \ion{O}{iii}] $\uplambda$1663 contribution from \ion{He}{ii} $\uplambda$1640.  As explained above \ion{O}{iii}] $\uplambda$1663 will be much less variable than \ion{He}{ii} and will not follow rapid variations of the continuum the way \ion{He}{ii} does.  For display purposes, to again emphasize that what matters is the slope, we have subtracted off a constant flux of $500 \times 10^{-15}$ ergs s$^{-1}$ cm$^{-2}$ from the reported \ion{He}{ii} $\uplambda$1640 fluxes to make the OLS-bisector line go through the origin.  These points are shown as filled circles in Figure 4.  Again, the subtraction of a constant has no effect on our reddening estimate which depends only on the slope of the line.  The slope of the OLS-bisector fit gives a mean \ion{He}{ii} $\uplambda$1640/$\uplambda$4686 ratio of $3.44 \pm 0.055$ and hence $E(\uplambda1640-\uplambda4686)$ of $0.98 \pm 0.24$. Our method thus gives a similar mean reddening for the 2013 {\it AGNSTORM} campaign as for the 1988-1989 observing campaign even though the \ion{He}{ii} flux levels were higher in 2013.

\begin{figure} 
	\label{fig:fig:HeIIAGNSTORMVectors}
	\begin{center}
		\includegraphics[width = 1\linewidth]{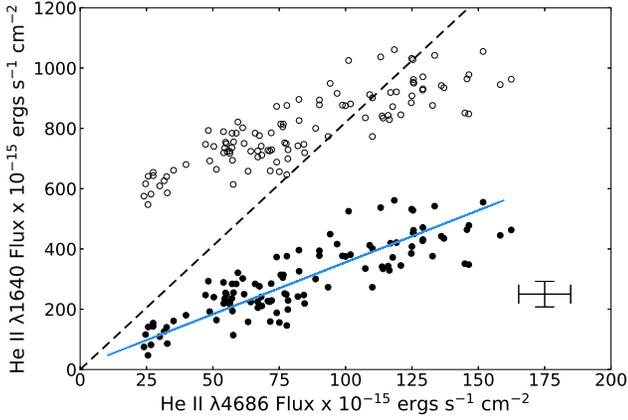}
        \caption{Near simultaneous \ion{He}{ii} $\uplambda$1640 vs \ion{He}{ii} $\uplambda$4686 fluxes from the 2013 {\it AGNSTORM} campaign. As in Figure 3, the dashed line is a theoretical \ion{He}{ii} $\uplambda$1640/$\uplambda$4686 ratio of 8.5.  The reported fluxes are shown as open circles.  The solid circles are the result of subtracting $500 \times 10^{-15}$ ergs s$^{-1}$ cm$^{-2}$ from the reported \ion{He}{ii} $\uplambda$1640 fluxes (see text). The solid (blue) line is an OLS-bisector fit to the shifted data.}
	\end{center}
\end{figure}

In summary, our two different methods of estimating the reddening of NGC~5548 from the \ion{He}{ii} lines for observing campaigns in three different years are consistent with an unweighted mean $E(\uplambda1640 - \uplambda4686) \thickapprox 0.94 \pm 0.10$ magnitudes.

\subsection{OI $\uplambda$1304/$\uplambda$8446}

\begin{figure}
 \label{fig:OI}
 \centering \includegraphics[width=8.5cm]{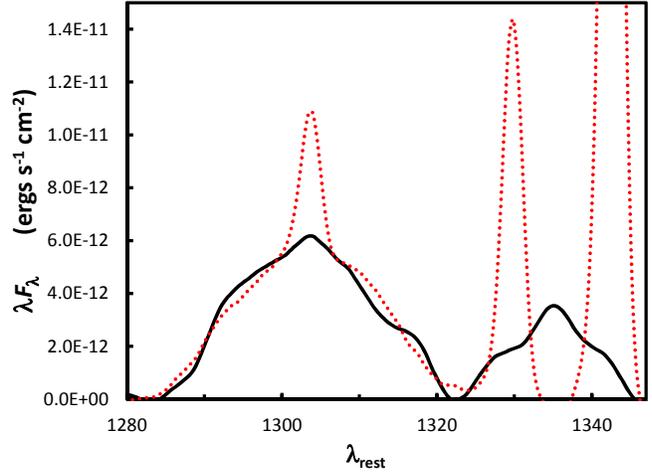}
 \caption{A continuum subtracted and smoothed average {\it HST} spectrum (solid black line) from the 1993 {\it International AGN Watch} campaign showing the region of \ion{O}{i} $\uplambda$1304 and \ion{C}{ii} $\uplambda$1335 emission with narrow absorption lines removed.  The dotted (red) line shows a smoothed blueshifted H$\upbeta$ profile scaled by a factor of 0.22 to \ion{O}{i} $\uplambda1304$.  Note that neither the narrow-line region contribution to H$\upbeta$ nor the [\ion{O}{iii}] $\uplambda$4959 $\uplambda$5007 lines have been removed.}
\end{figure}

Figure 5 shows the average \ion{O}{i} $\uplambda$1304 profile for NGC~5548 during the 1993 {\it HST} monitoring.  The two narrow absorption lines in the \ion{O}{i} profile (see Fig.\@ 4b of \citealt{Korista+95}) have been removed, the continuum subtracted, and the spectrum smoothed by a boxcar of half width 1.75 \AA.  A smoothed H$\upbeta$ profile taken from the simultaneous optical monitoring of \citet{Korista+95} is shown for comparison. This has been scaled by a factor of 0.22.  \citet{Grandi80} showed that for NGC~4151, \ion{O}{i} $\uplambda$8446 has the same profile as broad H$\upalpha$ implying that it comes only from the BLR.  It can be seen in Figure 5 that the \ion{O}{i} $\uplambda$1304 profile agrees well with the broad H$\upbeta$ profile and clearly lacks a narrow-line region (NLR) contribution.  This agreement with the H$\upbeta$ profile is important because there is no sign of \ion{Si}{ii} $\uplambda$1307 emission in the red wing of $\uplambda$1304.  The weaker line on the right-hand side of Figure 5 is the \ion{C}{ii}] $\uplambda$1335 line.  Thus, contrary to some suggestions (see Section 3.2),  \ion{Si}{ii} is not a significant contributor to the $\uplambda$1304 blend in NGC~5548.  

To our knowledge, \ion{O}{i} $\uplambda$8446 was not observed during the {\it HST} monitoring but it has been observed at other times. \citet{Landt+08} give average fluxes for \ion{O}{i} $\uplambda$8446 and H$\upbeta$ from high-quality spectra obtained on three nights in May 2004, January 2006, and June 2006.  These give an \ion{O}{i} $\uplambda$8446/H$\upbeta$ ratio of = 0.16. Earlier but lower-quality observations of \citet{Grandi80} gave a similar ratio of 0.18.  The relative strength of \ion{O}{i} $\uplambda$8446 compared with H$\upbeta$ in NGC~5548 is very close to the median relative strength in other AGNs (see Table 1 of \citealt{Grandi80}).  The \ion{O}{i} $\uplambda$8446/H$\upbeta$ ratio of \citet{Landt+08} and the \ion{O}{i} $\uplambda$1304/H$\upbeta$ ratio during the 1993 monitoring (see Figure 5) give an \ion{O}{i} $\uplambda$1304/$\uplambda$8446 ratio of $1.375 \pm 0.08$.  This gives $E(\uplambda1304-\uplambda8446) = 1.68 \pm 0.13$ magnitudes.

\subsection{Optical continuum colour (Cho{\l}oniewski method)}

\citet{Cackett+07} get a total reddening of $E(B-V) = 0.18$ by applying the Cho{\l}oniewski method to the $2003 - 2004$ photometry of \citet{Sergeev+05} and assuming that the unreddened colour of the variable component is given by a $F_{\nu} \propto \nu^{+1/3}$ power law.  Similarly, \citet{Sakata+10}, from their photometry obtained over the period $2001 - 2007$, get $E(B-V) = 0.13$.  

However, as discussed in Section 3.3.1, because of the higher-order Balmer lines and Balmer continuum in the $B$ filter, the unreddened slope in an $F_B$ vs.\@ $F_V$ plot will {\em not} be 1.07 as predicted by Eq.\@ 3.  The \citet{Cackett+07} and \citet{Sakata+10} reddening estimates are thus lower limits.  Using an unreddened gradient of 1.18 instead gives $E(B-V) \thickapprox 0.26 \pm 0.06$.

\subsection{Reddening from the UV to optical spectral slope}

The published UV/optical flux ratios plotted as grey circles in Figure 1 do not include allowance for the uncertain contribution of the host galaxy starlight to the optical continuum flux.  For NGC~5548 the starlight contribution to the optical flux {\em is} known \citep{Bentz+13}. We therefore plotted NGC~5548 in Figure 1 twice, once using the uncorrected optical flux (the filled circle) to match the uncorrected ratios for the other AGNs, and once with the starlight subtracted (the filled square). We have taken average values of the two ratios from \citet{Clavel+91} and \citet{Peterson+91} from the 1988-1989 {\it International AGN Watch} campaign. The uncertainty in the starlight correction for the AGNs in general is the main uncertainty in using the UV/optical continuum ratio to estimate reddenings.  The separation between the two NGC~5548 points in Figure 1 gives an indication of the errors. From Figure 1, $E(1337-5100) = 1.09 \pm 0.26$ where the uncertainty is the uncertainty in the host galaxy starlight subtraction.  It is interesting that if the points for the other AGNs (the grey circles) are shifted to the right because of an approximately similar starlight correction as for NGC~5548, the start of the reddening vector is at $\log F_{1337}/F_{5100} \sim 0.2$ which is the ratio predicted if an $F_{\nu} \propto \nu^{1/3}$ spectrum continues to $\sim 1300$.


\subsection{Lyman $\upalpha$ / H$\upbeta$}
%

Figure 1 shows the mean ratio of Ly$\upalpha$/H$\upbeta$ using average values from \citet{Clavel+91} and \citet{Peterson+91} the 1988-1989 {\it International AGN Watch} campaign, that had similar or the same Julian Dates. The average is 8.75.  The average error for individual ratios is $\pm 0.37$. This corresponds to $E(Ly\alpha-H\beta) = 1.50 \pm 0.06$.

\subsection{The Balmer decrement}
%

In the optical, the emphasis of the major multi-wavelength NGC~5548 reverberation mapping campaigns has been on H$\upbeta$ \citep{Peterson+91,Peterson+92,Korista+95}.  Nevertheless, many H$\upalpha$ flux measurements are also available (\citealt{Dietrich+93}). The average H$\upalpha$/H$\upbeta$ ratio is 3.70 for the 1988-1989 {\it International AGN Watch} campaign.  For an intrinsic ratio of 2.8 this gives $E(H\alpha-H\beta) = 0.30$ during the campaign. If we adopt an uncertainty in the intrinsic ratio of $\pm 0.1$, the error in $E(H\alpha-H\beta)$ due to this is $\pm 0.04$.  This does not include probable systematic errors in the measurement of the line ratio.

\subsection{Paschen $\upbeta$ / H$\upbeta$}
%

 From optical and IR observations made in 2004 and 2006 \citet{Landt+08} get a Pa$\upbeta$ flux of $1.63 \times 10^{-13}$ ergs s$^{-1}$ cm$^{-2}$ and a H$\upbeta$ flux of $3.92 \times 10^{-13}$ ergs s$^{-1}$ cm$^{-2}$.  These give a Pa$\upbeta$/H$\upbeta$ ratio of $0.42 \pm 0.6$.  These observations were not simultaneous.  Also they were made well outside the times of the UV-optical monitoring campaigns so there is a possibility that the reddening changed, but we include these  observations of the Pa$\beta$/H$\upbeta$ here because they provide additional evidence for large internal extinction in NGC~5548 consistent with other reddening indicators.

\begin{table*}
\begin{minipage}{126mm}
\caption{Reddening estimates for NGC~5548}
\label{symbols}
\begin{tabular}{@{}lcccccc}
\hline																	
Feature(s)	&	$\uplambda_{\rmn{short}}$	&	$\uplambda_{\rmn{long}}$	&	$E(B-V)$			&	$E(B-V)$			&	$E(B-V)$			\\
	&		&		&	Milky Way			&	Gaskell \& Benker			&	SMC			\\
\hline																	
Ly$\upalpha$/H$\upbeta$	&	$\uplambda$1216	&	$\uplambda$4861	&	0.22	$\pm$	0.02	&	0.31	$\pm$	0.03	&	0.09	$\pm$	0.01	\\
\ion{O}{i}	&	$\uplambda$1304	&	$\uplambda$8446	&	0.22	$\pm$	0.03	&	0.25	$\pm$	0.03	&	0.10	$\pm$	0.01	\\
$F_{\uplambda1337}$/$F_{\uplambda5100}$	&	$\uplambda$1350	&	$\uplambda$5100	&	0.19	$\pm$	0.05	&	0.22	$\pm$	0.05	&	0.07	$\pm$	0.02	\\
\ion{He}{ii}	&	$\uplambda$1640	&	$\uplambda$4686	&	0.25	$\pm$	0.02	&	0.24	$\pm$	0.02	&	0.10	$\pm$	0.01	\\
$F_B/F_V$	&	$\uplambda$4400	&	$\uplambda$5500	&	0.26	$\pm$	0.06	&	0.26	$\pm$	0.06	&	0.26	$\pm$	0.06	\\
H$\upalpha$/H$\upbeta$	&	$\uplambda$4861	&	$\uplambda$6563	&	0.27	$\pm$	0.03	&	0.27	$\pm$	0.03	&	0.28	$\pm$	0.03	\\
Pa$\upbeta$/H$\upbeta$	&	$\uplambda$4861	&	1.282 $\mu$m	&	0.25	$\pm$	0.12	&	0.25	$\pm$	0.12	&	0.26	$\pm$	0.12	\\
	&		&		&				&				&				\\
Mean	&		&		&	0.24	$\pm$	0.01	&	0.26	$\pm$	0.01	&	0.17	$\pm$	0.04	\\
\hline																				
\end{tabular}
\end{minipage}
\end{table*}

\section{Results for different reddening curves}

In Table 2 we summarize our reddening estimates for the seven indicators and the three reddening curves.  As in Table 1, we have ordered the rows by increasing average wavelength of the indicators.  Several things are immediately apparent from Table 2.  
\begin{enumerate}[1.]
\item {\em All} reddening estimates give reddenings that are {\em far} greater than the small $E(B-V) = 0.018$ reddening due to dust in the solar neighbourhood.  This is true for {\em each} of the reddening curves considered.
\item The derived {\em mean} reddening does not depend strongly on the choice of reddening curve.
\item  The SMC reddening curve produces $E(B-V)$ estimates from the UV that are incompatible with estimates from longer wavelength reddening indicators.
\item For the Gaskell \& Benker and the Milky Way reddening curves, the dispersion of the individual reddening estimates about the mean is only slightly larger than the errors in the individual estimates.
\item  For the Gaskell \& Benker and the Milky Way reddening curves, the hydrogen-line reddening indicators give similar reddenings to the non-hydrogen-line indicators.
\end{enumerate}

Additional support for the \citet{Gaskell+Benker07} reddening curve being appropriate for NGC~5548 is that comparison of the SED of NGC~5548 with a bluer AGN, such as 3C~273, gives a reddening curve similar to the Gaskell \& Benker curve (see Figure 5 of \citealt{Gaskell+07}).

\section{The Luminosity of NGC~5548}

It has long been realized \citep{Wampler68} that if there is substantial reddening of the continuum of AGNs, then the UV luminosity will be much higher.  Because we are using reddening indicators that go quite far into the UV and are hence measuring the selective extinction between the optical and the UV directly, the total extinction in the UV is well determined and this does {\em not} depend on the choice of reddening curve adopted.  For example, if we take the reddening implied by the \ion{O}{i} lines (see Table 2), the Milky Way, Gaskell \& Benker. and SMC reddening curves give very similar attenuations at $\uplambda$1200 of factors of eight, seven, and seven respectively.  We can thus conclude that the real luminosity at $\uplambda$1200 is around a factor of seven greater than observed.  Since the reddening curve is probably flat at shorter wavelengths, this means that the bolometric luminosity is also a factor of seven greater than observed.

\section{The size of the accretion disc}

Because they are so physically thick, accretion discs in AGNs are also optically thick.  Accretion disks have similar temperature structures.  Except at the very shortest wavelengths, the luminosity of an accretion disc at a given wavelength depends on its surface area and hence on the square of the effective radius.  This readily permits one to calculate what the size of the region emitting a given wavelength will be. Continuum reverberation mapping of 14 AGNs by \citet{Sergeev+05} indicated that the sizes of AGN accretion discs were larger than predicted.  Using the very different technique of microlensing, \citet{Pooley+06,Pooley+07} found that the inferred sizes of the accretion discs in 10 AGNs were also too large.  This is referred to as the ``accretion disc size problem". If it is a real discrepancy, it means that something is seriously wrong with our standard optically-thick accretion disc model for AGNs. 

\citet{Sergeev+05} pointed out that the accretion disc size discrepancy could be solved if the luminosities of the AGNs had simply been underestimated.  \citet{Gaskell17} proposed that this was indeed the solution to the disc-size problem and that the luminosities of AGNs were being underestimated by an order of magnitude by neglecting internal reddening.  We can quantitatively test this with NGC~5548.   For NGC~5548, \citet{Edelson+15} find the size of the accretion disc to be 2.5 times larger than predicted (see their section 4.1).  This is similar to the typical size discrepancy in micro-lensed AGNs and other reverberation-mapped AGNs (see Figure 6 of \citealt{Edelson+15}).

The bulk of the luminosity of a thermal AGN such as NGC~5548 comes out at wavelengths shorter the $\uplambda$1200 (see Figure 2 of \citealt{Gaskell08}). Since including a correction for internal reddening makes the luminosity of NGC~5548 $\thickapprox 7$ times greater (see previous section), this makes the size of the accretion disc $\sqrt{7} = 2.6$ times bigger.  This is thus in good agreement with the larger disc size found by \citealt{Edelson+15}) and hence supports the hypothesis that the disc-size problem is simply due to the neglect of internal reddening.

\section{Discussion}

\subsection{Hydrogen line ratios}

An important point is that {\em the seven reddening indicators we consider here give consistent reddening values for NGC~5548.}  Although there can be practical problems in determining the quantities going into each individual indicator, and although in some cases the observations were not simultaneous, the agreement in reddening estimates in Table 2 shows that, contrary to previous concerns, there are no major problems with the fundamental intrinsic ratios for these indicators. In particular, this includes the hydrogen lines.  We can see from Table 2 that the mean reddening from the three hydrogen line ratios is the same as the mean from the four non-hydrogenic indicators.  We thus conclude that what has long been referred to as ``the hydrogen-line problem" (Lyman $\upalpha$ being too weak relative to the Balmer lines, and the Balmer decrement being too steep) is not a real problem; it is just a consequence of greatly underestimating the reddening. The same goes for the so-called ``Balmer line crisis" (the Balmer lines being too strong relative to the UV lines).  The hydrogen line ratios are close to their Case B values.  \citet{Gaskell17} argues that the reason for this that real BLR clouds are unlike clouds modelled in photoionization codes.  The latter are high-column-density clouds which emit only out of the front of the cloud (towards the source of ionizing radiation) or out of the back.  Real BLR clouds have lower column densities and readily emit sideways (see Figure 3 of \citealt{Gaskell17}).  This means that optical depths are relatively small.

\subsection{Reprocessing by dust}

Energy absorbed by dust from the far UV (FUV) where AGN SEDs peak has to be re-radiated by the dust.  This issue has been addressed by \citet{Carleton+87} who considered the overall SEDs of AGNs from hard X-rays to the far red and considered similar reddenings to what we find for NGC~5548.  Their Figure 8 is particularly useful.  It shows SEDs plotted on a linear $\nu F_{\nu}$ scale against $\log \nu$, a plot which has the virtue that areas under curves represent energy. From their comparison of the SEDs of AGNs with low reddening (their Class A) with the SEDs of AGNs with reddenings similar to NGC~5548 (their Class B), it can readily be appreciated that the energy lost in the UV in Class B objects compared with Class A objects is reradiated in the IR from 3 -- 100 $\upmu$m.  There is thus no energetics problem with having large reddenings of AGNs.

\subsection{Variable reddening?}

If the main factor determining the H$\upalpha$/H$\upbeta$ ratio is the reddening -- as we are advocating here -- then changes in the ratio imply changes in the reddening. This was proposed by \citet{Goodrich89} as a cause of what are now referred to as ``changing-look" AGNs. In the present paper we are focusing on validating reddening indicators and estimating a mean reddening of NGC~5548.  We will therefore defer discussion of evidence for variable reddening for the future.  We note, however, that \citet{Shapovalova+04} present evidence for H$\upalpha$/H$\upbeta$ varying in NGC 5548 both over time and with velocity.  \citet{Gaskell+Harrington18} show this is consistent with variable partial obscuration of the BLR -- something supported by absorption line and X-ray variability.  Given the possibility of variations in the reddening, the mean reddening we find here might not be applicable for other other years.

\subsection{The broad-line region energy budget}

\citet{Wampler68} remarked that ``if the continuum is not reddened ... extrapolation of the visual continuum into the ultraviolet indicates that in this case there would be insufficient ultraviolet quanta to ionize the gas."  \citet{MacAlpine81} showed that there was a severe AGN ``energy-budget" problem because observed continua needed an impossible covering factor of $\Omega/4\uppi \gtrsim 9$ to match observed \ion{He}{ii} $\uplambda$4686 equivalent widths (see \citealt{MacAlpine03} for a review).  \citet{Gaskell+07} examined the energy-budget problem for NGC~5548 for other broad lines and concluded that the problem could only be solved by with significant internal reddening.  The reddening we find here is consistent with what the modelling of \citet{Gaskell+07} showed was needed.  A very low reddening correcting, such as using only the $E(B-V) \sim 0.02$ due to Galactic reddening alone, fails to give enough ionizing photons to reproduce the observed broad emission lines in an AGN like NGC~5548.

\subsection{Black hole masses}

The estimated mass of the black hole in NGC~5548 and, by extension, the masses of other black holes, are not affected by reddening.  The mass of the NGC~5548 black hole is estimated using the square of the full width at half maximum (FWHM) of lines times an effective size, $R$, which is determined by reverberation mapping, divided by the gravitational constant, and then multiplied by a factor, $f$, that depends on the geometry and kinematics of the BLR.  The factor $f$ is calibrated against black holes whose masses have been determined from spatially-resolved stellar dynamics.  None of this involves depends on the luminosity of the AGN.

\citet{Dibai77} proposed estimating black hole masses by estimating $R$ from the optical luminosity, $L_{opt}$, and assuming that $R = const \times L^{\alpha}$.  The scale factor and the slope, $\alpha$, of the power law of this relationship are calibrated against reverberation-mapped AGNs.  A systematic increase in the optical luminosity of AGNs after correcting for reddening merely changes the constant of proportionality.  Masses are not systematically increased.  Note that it should be possible to improve the $R-L$ relationship somewhat by allowing for the reddening of individual AGNs.

\subsection{Bolometric corrections and Eddington ratios}

In the absence of extensive multi-wavelength observations, bolometric luminosities of AGNs are estimated by multiplying a monochromatic luminosity in the optical or UV by a bolometric correction factor. Substantial reddening of AGNs means that the luminosity in the far UV has been underestimated (see above) and hence that bolometric corrections have been systematically underestimated by a significant factor. Higher bolometric corrections in turn increase estimates of Eddington ratios.

The energy lost at shorter wavelengths because of absorption by dust reappears in the infra-red (see Section 9.3).  In principle, therefore, a bolometric luminosity based on idealized observations of the {\em entire} electromagnetic spectrum including the IR will not be affected by absorption. The practical difficulty is in separating out emission in the mid- and far-IR from dust heated by the AGN from emission from dust dust heated by stars. Recognition of the importance of absorption at shorter wavelengths means that less of the mid-IR is due to heating by stars.

\subsection{The So{\l}tan argument}

\citet{Lynden-Bell69} pointed out that there should be supermassive black holes in every massive galaxy.  \citet{Soltan82} argued that the present-day masses of these black holes had to be consistent with the energy radiated by accretion onto the black holes over the history of the universe.  Although this argument is conceptually simple, obtaining the appropriate numbers is complicated.  In a detailed study of the So{\l}tan argument \citet{Marconi+04} find that an average Eddington ratio of 50\% is needed (see their Figure 7).  This favours higher Eddington ratios and supports the larger bolometric corrections implied by reddening corrections such as what we find for NGC~5548.

\subsection{Determining the reddening of AGNs}

We have shown that the reddening of NGC~5548, the best-studied type-1 thermal AGN, is substantial.  As far as line and continuum ratios go, NGC~5548 is quite typical. so the reddening of most thermal AGNs must also be substantial.  Reddening corrections {\em must} therefore be applied to an AGN first before trying to model SEDs.  With only slight modifications to observing campaigns, reddenings can be reliably determined.  The modifications are:
\begin{enumerate}[(1)]
    \item Observe H$\upalpha$ as well as H$\upbeta$
    \item Observe a broad enough wavelength region of the optical spectrum to be able to use the Cho{\l}oniewski method.
    \item If the observations include the UV, get simultaneous \ion{He}{ii} $\uplambda$1640 and $\uplambda$1640 observations for at least a relatively high and relatively low state.
    \item Again, if observing in the UV, Obtain a red spectrum including \ion{O}{i} $\uplambda$8446
    \item If possible, get a near-IR spectrum including Pa$\upbeta$
\end{enumerate}

Broad-band optical monitoring can also provide an additional reddening indicator using the Cho{\l}oniewski method.

\section{Conclusions}

Our conclusions are as follows:

\begin{enumerate}[(1)]

\item We have shown two methods by which reddenings can be obtained from \ion{He}{ii} lines: either by looking at optical and UV difference spectra, or by looking at the slope of the relationship between \ion{He}{ii} $\uplambda$1640 and $\uplambda$1640 fluxes as they vary.

\item \ion{Si}{ii} contamination of the \ion{O}{i} $\uplambda$1304 is negligible.
    
\item Seven separate reddening indicators point to NGC~5548 having a total reddening (Galactic + internal) of $E(B-V) \thickapprox 0.26 \pm 0.02$, which is some fifteen times the Galactic reddening alone.

\item The unreddened spectral energy distribution of the continuum from the optical to $\thickapprox \uplambda1200$ is consistent with the expected $F_{\nu} \propto \nu^{+1/3}$ of an externally-illuminated accretion disc.

\item The velocity-integrated flux ratios of the broad hydrogen lines are close to Case B values.

\item A steep SMC-like reddening curve is ruled out for NGC~5548.

\item If the reddening of NGC~5548 is $E(B-V) = 0.26$, this gives an extinction in the $V$ band, $A_V = 0.8$ magnitudes, which is attenuation of just over a factor of two.

\item  The UV luminosity of NGC~5548 around $\uplambda$1200 has been underestimated by a factor of $\thickapprox 7$. This value only depends weakly on the choice of reddening curve.

\item  The higher luminosity predicts an accretion disc size that is $\thickapprox 2.6$ times larger than would otherwise be predicted.  This solves the ``accretion disc size problem".

\item  Higher luminosities in the far UV explain the higher IR emission in reddened AGNs.

\item Because radiation absorbed in the optical to far UV must be re-radiated in the IR, the contribution of AGN heating to the mid-IR emission of an AGN is greater than has been thought and the contribution of heating by hot stars correspondginly less.

\item Eddington ratios are higher and this helps reconcile AGN counts with local black hole demographics (the So{\l}tan argument).

\end{enumerate}

\section*{Acknowledgments}

FCA, SAB and SG carried out their work under the auspices of the Science Internship Program (SIP) of the University of California at Santa Cruz.  We wish to express our appreciation to Raja GuhaThakurta for his excellent leadership of the program. MG is grateful to Ski Antonnucci and Clio Heard for useful discussions.

\section*{Data availability}

Data from the {\it International AGN Watch} monitoring campaigns are available at https://www.asc.ohio-state.edu/astronomy/agnwatch/data.html. Other data are available in the references cited.

\label{lastpage}


\begin{thebibliography}{67}

\bibitem[\protect\citeauthoryear{Antonucci}{2012}]{Antonucci12} Antonucci, R.~R.~J. 2012, Astron. \& and Astrophys. Trans., 27, 557

\bibitem[\protect\citeauthoryear{Baker, Menzel, \& Aller}{1938}]{Baker+38} Baker J.~G., Menzel D.~H., Aller L.~H., 1938, ApJ, 88, 422. doi:10.1086/143997

\bibitem[\protect\citeauthoryear{Baldwin et al.}{1996}]{Baldwin+96} Baldwin, J.~A., Ferland, G.~J., Korista, K.~T., et al.\ 1996, ApJ, 461, 664

\bibitem[\protect\citeauthoryear{Bechtold et al.}{1997}]{Bechtold+97} Bechtold, J., Shields, J., Rieke, M., et al.\ 1997, IAU Colloq.~159: Emission Lines in Active Galaxies: New Methods and Techniques, 113, 122

\bibitem[\protect\citeauthoryear{Bentz et al.}{2013}]{Bentz+13} Bentz, M.~C., Denney, K.~D., Grier, C.~J., et al.\ 2013, ApJ, 767, 149


\bibitem[\protect\citeauthoryear{Bottorff et al.}{2002}]{Bottorff+02} Bottorff, M.~C., Baldwin, J.~A., Ferland, G.~J., Ferguson, J.~W.,
\& Korista, K.~T.\ 2002, ApJ, 581, 932

\bibitem[\protect\citeauthoryear{Cackett, Horne, \& Winkler}{2007}]{Cackett+07} Cackett E.~M., Horne K., Winkler H., 2007, MNRAS, 380, 669. doi:10.1111/j.1365-2966.2007.12098.x

\bibitem[\protect\citeauthoryear{Carleton et al.}{1987}]{Carleton+87} Carleton, N.~P., Elvis, M., Fabbiano, G., et al.\ 1987, ApJ, 318, 595

\bibitem[\protect\citeauthoryear{Chamberlain}{1953}]{Chamberlain53} Chamberlain J.~W., 1953, ApJ, 117, 387. doi:10.1086/145704

\bibitem[\protect\citeauthoryear{Cho{\l}oniewski}{1981}]{Choloniewski81} Cho{\l}oniewski, J.\ 1981, Act.~Astron.~Ap., 31, 293

\bibitem[\protect\citeauthoryear{Clavel et al.}{1991}]{Clavel+91} Clavel, J., et al.\ 1991, ApJ, 366, 64

\bibitem[\protect\citeauthoryear{Crenshaw}{1986}]{Crenshaw86} Crenshaw, D.~M.\ 1986, ApJS, 62, 821

\bibitem[\protect\citeauthoryear{Czerny et al.}{2004}]{Czerny+04} Czerny, B., Li, J., Loska, Z., \& Szczerba, R.\ 2004, MNRAS, 348, L54

\bibitem[\protect\citeauthoryear{De Rosa et al.}{2015}]{deRosa+15} De Rosa G., Peterson B.~M., Ely J., Kriss G.~A., Crenshaw D.~M., Horne K., Korista K.~T., et al., 2015, ApJ, 806, 128. doi:10.1088/0004-637X/806/1/128

\bibitem[\protect\citeauthoryear{De Rosa et al.}{2018}]{deRosa+18} De Rosa G., Fausnaugh M.~M., Grier C.~J., Peterson B.~M., Denney K.~D., Horne K., Bentz M.~C., et al., 2018, ApJ, 866, 133. doi:10.3847/1538-4357/aadd11

\bibitem[\protect\citeauthoryear{Dibai}{1977}]{Dibai77} Dibai, E.~A.\ 1977, Soviet Astron. Lett., 3, 1

\bibitem[\protect\citeauthoryear{Dietrich et al.}{1993}]{Dietrich+93} Dietrich M., Kollatschny W., Peterson B.~M., Bechtold J., Bertram R., Bochkarev N.~G., Boroson T.~A., et al., 1993, ApJ, 408, 416. doi:10.1086/172599 

\bibitem[\protect\citeauthoryear{Dietrich et al.}{2001}]{Dietrich+01} Dietrich, M., Bender, C.~F., Bergmann, D.~J., et al.\ 2001, A\&Ap, 371, 79

\bibitem[\protect\citeauthoryear{Edelson et al.}{2015}]{Edelson+15} Edelson R., Gelbord J.~M., Horne K., McHardy I.~M., Peterson B.~M., Ar{\'e}valo P., Breeveld A.~A., et al., 2015, ApJ, 806, 129. doi:10.1088/0004-637X/806/1/129

\bibitem[\protect\citeauthoryear{Fausnaugh et al.}{2016}]{Fausnaugh+16} Fausnaugh M.~M., Denney K.~D., Barth A.~J., Bentz M.~C., Bottorff M.~C., Carini M.~T., Croxall K.~V., et al., 2016, ApJ, 821, 56. doi:10.3847/0004-637X/821/1/56

\bibitem[\protect\citeauthoryear{Ferland}{1999}]{Ferland99} Ferland G.~J., 1999, PASP, 111, 1524. doi:10.1086/316466

\bibitem[\protect\citeauthoryear{Ferland \& Shields}{1985}]{Ferland+Shields85} Ferland G.~J., Shields G.~A., 1985, in Astrophysics of active galaxies and quasi-stellar objects (Mill Valley, CA, University Science Books), pp. 157-184.

\bibitem[\protect\citeauthoryear{Friedjung}{1985}]{Friedjung85} Friedjung, M.\ 1985, A\&Ap, 146, 366

\bibitem[\protect\citeauthoryear{Gaskell}{1982}]{Gaskell82} Gaskell, C.~M.\ 1982, PASP, 94, 891

\bibitem[\protect\citeauthoryear{Gaskell}{1984}]{Gaskell84} Gaskell, C.~M.\ 1984, ApLett, 24, 43

\bibitem[\protect\citeauthoryear{Gaskell}{2007}]{Gaskell07} Gaskell C.~M., 2007, ASPC, 373, 596

\bibitem[\protect\citeauthoryear{Gaskell}{2009}]{Gaskell09} Gaskell, C.~M. 2009, New Astron. Rev, 53, 140

\bibitem[\protect\citeauthoryear{Gaskell}{2008}]{Gaskell08} Gaskell, C.~M.\ 2008, Revista Mexicana de Astron.\@ y Astrofis.\@ Conf.\@ Ser., 32, 1

\bibitem[\protect\citeauthoryear{Gaskell}{2017}]{Gaskell17} Gaskell, C.~M., 2017, MNRAS, 467, 226. doi:10.1093/mnras/stx094

\bibitem[\protect\citeauthoryear{Gaskell et al.}{2021}]{Gaskell+21} Gaskell C.~M., Bartel K., Deffner J.~N., Xia I., 2021, MNRAS, 508, 6077. doi:10.1093/mnras/stab2443

\bibitem[\protect\citeauthoryear{Gaskell \& Benker}{2007}]{Gaskell+Benker07} Gaskell, C.~M., \& Benker, A.~J.\ 2007, arXiv:0711.1013

\bibitem[\protect\citeauthoryear{Gaskell, Gill, \& Singh}{2016}]{Gaskell+16} Gaskell C.~M., Gill J.~J.~M., Singh J., 2016, arXiv, arXiv:1611.03733

\bibitem[\protect\citeauthoryear{Gaskell et al.}{2004}]{Gaskell+04} Gaskell, C.~M., Goosmann, R.~W., Antonucci, R.~R.~J.,
\& Whysong, D.~H.\ 2004, ApJ, 616, 147

\bibitem[\protect\citeauthoryear{Gaskell \& Harrington}{2018}]{Gaskell+Harrington18} Gaskell C.~M., Harrington P.~Z., 2018, MNRAS, 478, 1660. doi:10.1093/mnras/sty848

\bibitem[\protect\citeauthoryear{Gaskell et al.}{2007}]{Gaskell+07} Gaskell, C.~M., Klimek, E.~S., \&
Nazarova, L.~S.\ 2007, ApJ, submitted [arXiv:0711.1025]

\bibitem[\protect\citeauthoryear{Gaskell et al.}{2022}]{Gaskell+22} Gaskell M., Thakur N., Tian B., Saravanan A., 2022, AN, 343, e210112. doi:10.1002/asna.20210112

\bibitem[\protect\citeauthoryear{Goodrich}{1989}]{Goodrich89} Goodrich R.~W., 1989, ApJ, 340, 190. doi:10.1086/167384

\bibitem[\protect\citeauthoryear{Grandi}{1980}]{Grandi80} Grandi, S.~A.\ 1980, ApJ, 238, 10

\bibitem[\protect\citeauthoryear{Grandi}{1983}]{Grandi83} Grandi, S.~A.\ 1983, ApJ, 268, 591

\bibitem[\protect\citeauthoryear{Heard \& Gaskell}{2022}]{Heard+Gaskell22} Heard C.~Z.~P., Gaskell C.~M., 2022, MNRAS in press. doi:10.1093/mnras/stac2220 [arXiv, arXiv:2108.09995]

\bibitem[\protect\citeauthoryear{Isobe}{1990}]{Isobe+90} Isobe, T., Feigelson, E.~D., Akritas, M.~G., \& Babu, G.~J.\ 1990, ApJ, 364, 104

\bibitem[\protect\citeauthoryear{Kaspi et al.}{2000}]{Kaspi+00} Kaspi S., Smith P.~S., Netzer H., Maoz D., Jannuzi B.~T., Giveon U., 2000, ApJ, 533, 631. doi:10.1086/308704

\bibitem[\protect\citeauthoryear{Kelly et al.}{2010}]{Kelly+10} Kelly B.~C., Vestergaard M., Fan X., Hopkins P., Hernquist L., Siemiginowska A., 2010, ApJ, 719, 1315. doi:10.1088/0004-637X/719/2/1315

\bibitem[\protect\citeauthoryear{Kollmeier et al.}{2006}]{Kollmeier+06} Kollmeier J.~A., Onken C.~A., Kochanek C.~S., Gould A., Weinberg D.~H., Dietrich M., Cool R., et al., 2006, ApJ, 648, 128. doi:10.1086/505646

\bibitem[\protect\citeauthoryear{Koratkar \& Gaskell}{1991}]{Koratkar+Gaskell91} Koratkar A.~P., Gaskell C.~M., 1991, ApJL, 370, L61. doi:10.1086/185977

\bibitem[\protect\citeauthoryear{Korista et al.}{1995}]{Korista+95} Korista, K.~T., et al.\ 1995, ApJS, 97, 285

\bibitem[\protect\citeauthoryear{Kriss et al.}{2019}]{Kriss+19} Kriss G.~A., De Rosa G., Ely J., Peterson B.~M., Kaastra J., Mehdipour M., Ferland G.~J., et al., 2019, ApJ, 881, 153. doi:10.3847/1538-4357/ab3049

\bibitem[\protect\citeauthoryear{Krolik et al.}{1991}]{Krolik+91} Krolik J.~H., Horne K., Kallman T.~R., Malkan M.~A., Edelson R.~A., Kriss G.~A., 1991, ApJ, 371, 541. doi:10.1086/169918

\bibitem[\protect\citeauthoryear{Kwan \& Krolik}{1981}]{Kwan+Krolik81} Kwan, J., \& Krolik, J.~H.\ 1981, ApJ, 250, 478

\bibitem[\protect\citeauthoryear{Landt et al.}{2008}]{Landt+08} Landt, H., Bentz, M.~C.,
Ward, M.~J., et al.\ 2008, ApJS, 174, 282

\bibitem[\protect\citeauthoryear{Lynden-Bell}{1969}]{Lynden-Bell69}  Lynden-Bell, D.\ 1969, Nature, 223, 690

\bibitem[\protect\citeauthoryear{MacAlpine}{1981}]{MacAlpine81} MacAlpine, G.~M.\ 1981, ApJ, 251, 465

\bibitem[\protect\citeauthoryear{MacAlpine}{2003}]{MacAlpine03} MacAlpine G.~M., 2003, RMxAC, 18, 63

\bibitem[\protect\citeauthoryear{Marconi et al.}{2004}]{Marconi+04} Marconi A., Risaliti G., Gilli R., Hunt L.~K., Maiolino R., Salvati M., 2004, MNRAS, 351, 169. doi:10.1111/j.1365-2966.2004.07765.x

\bibitem[\protect\citeauthoryear{Mathews \& Wampler}{1985}]{Mathews+Wampler85} Mathews, W.~G., \& Wampler, E.~J.\ 1985, PASP, 97, 966

\bibitem[\protect\citeauthoryear{Matsuoka et al.}{2007}]{Matsuoka+07} Matsuoka, Y., Oyabu, S., Tsuzuki, Y., \& Kawara, K.\ 2007, ApJ, 663, 781

\bibitem[\protect\citeauthoryear{Mehdipour et al.}{2015}]{Mehdipour+15} Mehdipour M., Kaastra J.~S., Kriss G.~A., Cappi M., Petrucci P.-O., Steenbrugge K.~C., Arav N., et al., 2015, A\&A, 575, A22. doi:10.1051/0004-6361/201425373

\bibitem[\protect\citeauthoryear{Netzer \& Davidson}{1979}]{Netzer+Davidson79} Netzer, H., \& Davidson, K.\ 1979, MNRAS, 187, 871

\bibitem[\protect\citeauthoryear{Netzer et al.}{1995}]{Netzer+95} Netzer, H., Brotherton, M.~S., Wills, B.~J., et al.\ 1995, ApJ, 448, 27


\bibitem[\protect\citeauthoryear{Osterbrock}{1977}]{Osterbrock77} Osterbrock, D.~E.\ 1977, ApJ, 215, 733

\bibitem[\protect\citeauthoryear{Osterbrock \& Ferland}{2006}]{Osterbrock+Ferland06} Osterbrock, D.~E., \& Ferland, G.~J.\ 2006, Astrophysics of Gaseous Nebulae and Active Galactic Nuclei, 2nd.~ed.~by D.E.~Osterbrock and G.J.~Ferland.~Sausalito, CA: University Science Books

\bibitem[\protect\citeauthoryear{Padovani \& Rafanelli}{1988}]{Padovani+Rafanelli88} Padovani P., Rafanelli P., 1988, A\&A, 205, 53

\bibitem[\protect\citeauthoryear{Pei et al.}{2017}]{Pei+17} Pei L., Fausnaugh M.~M., Barth A.~J., Peterson B.~M., Bentz M.~C., De Rosa G., Denney K.~D., et al., 2017, ApJ, 837, 131. doi:10.3847/1538-4357/aa5eb1

\bibitem[\protect\citeauthoryear{Peterson et al.}{1991}]{Peterson+91} Peterson, B.~M., Balonek, T.~J., Barker, E.~S., et al.\ 1991, ApJ, 368, 119

\bibitem[\protect\citeauthoryear{Peterson et al.}{1992}]{Peterson+92} Peterson, B.~M., Alloin, D., Axon, D., et al.\ 1992, ApJ, 392, 470

\bibitem[\protect\citeauthoryear{Pooley et al.}{2006}]{Pooley+06} Pooley, D., Blackburne, J.~A., Rappaport, S., Schechter, P.~L., \& Fong, W.-f.\ 2006, ApJ, 648, 67

\bibitem[\protect\citeauthoryear{Pooley et al.}{2007}]{Pooley+07} Pooley, D., Blackburne, J.~A., Rappaport, S., \& Schechter, P.~L.\ 2007, ApJ, 661, 19

\bibitem[\protect\citeauthoryear{Richards et al.}{2003}]{Richards+03} Richards, G.~T., Hall, P.~B., Vanden Berk, D.~E., et al.\ 2003, AJ, 126, 1131

\bibitem[\protect\citeauthoryear{Rodr{\'{\i}}guez-Ardila et al.}{2002}]{Rodriguez-Ardila+02} Rodr{\'{\i}}guez-Ardila, A., Viegas, S.~M., Pastoriza, M.~G., Prato, L., \& Donzelli, C.~J.\ 2002, ApJ, 572, 94

\bibitem[\protect\citeauthoryear{Sakata et al.}{2010}]{Sakata+10} Sakata, Y., Minezaki, T., Yoshii, Y., et al.\ 2010, ApJ, 711, 461

\bibitem[\protect\citeauthoryear{Schlafly \& Finkbeiner}{2011}]{Schlafly+Finkbeiner11} Schlafly, E.~F., \& Finkbeiner, D.~P.\ 2011, ApJ, 737, 103

\bibitem[\protect\citeauthoryear{Schlegel et al.}{1998}]{Schlegel+98} Schlegel, D.~J., Finkbeiner, D.~P., \& Davis, M.\ 1998, ApJ, 500, 525

\bibitem[\protect\citeauthoryear{Sergeev et al.}{2005}]{Sergeev+05} Sergeev, S.~G., Doroshenko, V.~T., Golubinskiy, Yu.~V., Merkulova, N.~I., \& Sergeeva, E.~A.\ 2005, ApJ, 622, 129

\bibitem[\protect\citeauthoryear{Shapovalova et al.}{2004}]{Shapovalova+04} Shapovalova, A.~I., Doroshenko, V.~T., Bochkarev, N.~G., et al.\ 2004, A\&Ap, 422, 925

\bibitem[\protect\citeauthoryear{Shuder}{1982}]{Shuder82} Shuder, J.~M.\ 1982, ApJ, 259, 48

\bibitem[\protect\citeauthoryear{Shuder \& MacAlpine}{1979}]{Shuder+MacAlpine79} Shuder, J.~M., \& MacAlpine, G.~M.\ 1979, ApJ, 230, 348

\bibitem[\protect\citeauthoryear{So{\l}tan}{1982}]{Soltan82} So{\l}tan, A.\ 1982, MNRAS, 200, 115


\bibitem[\protect\citeauthoryear{Wampler}{1967}]{Wampler67} Wampler, E.~J.\ 1967, PASP, 79, 210

\bibitem[\protect\citeauthoryear{Wampler}{1968}]{Wampler68} Wampler, E.~J.\ 1968, AJ, 73, 855

K.~D., Vestergaard, M., \& Davis, T.~M.\ 2011, ApJL, 740, L49

\bibitem[\protect\citeauthoryear{Weingartner \& Draine}{2001}]{Weingartner+Draine01} Weingartner, J.~C., \& Draine, B.~T.\ 2001, ApJ, 548, 296

\bibitem[\protect\citeauthoryear{Winkler}{1997}]{Winkler97} Winkler, H.\ 1997, MNRAS,
292, 273

\bibitem[\protect\citeauthoryear{Winkler et al.}{1992}]{Winkler+92} Winkler, H., Glass, I.~S., van Wyk, F., et al.\ 1992, MNRAS, 257, 659

\bibitem[\protect\citeauthoryear{Wysota \& Gaskell}{1988}]{Wysota+Gaskell88} Wysota, A., \& Gaskell, C.~M.\ 1988, in Active Galactic Nuclei, ed. H.~R. Miller \& P.~J. Wiita, Lecture Notes in Physics, 307, 79

\bibitem[\protect\citeauthoryear{Zheng}{1992}]{Zheng92} Zheng, W.\ 1992, ApJ, 385, 127

\bibitem[\protect\citeauthoryear{}{}]{}

\end{thebibliography}
\end{document}